\documentstyle[preprint,prd,aps,epsf,epsfig]{revtex}
\begin{document}
\draft
%
\title{ Self-organization in the Concert Hall: the Dynamics of Rhythmic
Applause}


\maketitle
\medskip

An audience expresses appreciation for a good performance by the
strength and nature of its applause.  The initial thunder often
turns into synchronized clapping - an event familiar to many who
frequent concert halls.  Synchronized clapping has a well defined
scenario: the initial strong incoherent clapping is followed by a
relatively sudden synchronization process, after which everybody claps
simultaneously and periodically.  This synchronization can disappear
and reappear several times during the applause. The phenomenon is a
delightful expression of social self-organization, that provides a
human scale example of the synchronization processes observed in
numerous systems in nature ranging from the synchronized flashing of
the south-east Asian fireflies to oscillating chemical reactions
\cite{strogatz,bottani,glass}.

Here we investigate the mechanism and the development of synchronized
clapping by performing a series of measurements focused on both
the collective aspects of the self-organization process as well as
the behavior of the individuals in the audience.  We recorded several
theater and opera performances in Eastern Europe (Romania and
Hungary) utilizing a microphone placed at the ceiling of the hall
(Fig~1a).  Typically, after a few seconds of random clapping a
periodic signal develops (a signature of synchronized clapping),
clearly visible in Fig. 1a as pronounced pikes in the signal. This
transition is also captured by the order parameter (Fig.~1c), which
increases as the periodic signal develops, and decreases as it
disappears. While synchronization increases the strength of the signal
at the moment of the clapping, it leads to a decrease in the average
noise intensity in the room (see Fig.~1d).  This is rather
surprising, since one would expect that the driving force for
synchronization is the desire of the audience to express its
enthusiasm by increasing the average noise intensity.  The origin of
this conflict between the average noise and synchronization can be
understood by correlating the global signal with the behavior of an
individual in the audience.  For this we recorded the local sound
intensity in the vicinity of a group of individuals (Fig.~1b),
unaware of the recording process.  In the incoherent phase the local
signal is periodic with a short period corresponding to the fast
clapping of an individual in the audience.  However, the clapping
period suddenly doubles at the beginning of the synchronized phase
(approximately at 12$s$ in Fig. 1a and b), and slowly decreases as
synchronization is lost (Fig.~1e).  Thus, the decrease in the average
noise intensity is a consequence of the period doubling, since there
is less clapping in unit time.  An increase in the average noise
intensity is possible only by decreasing the clapping period, which
indeed does take place, as shown in Fig.~1e. However, the decreasing
clapping period gradually brings the synchronized clapping back to
the fast clapping observed in the early asynchronous phase, and
synchronization disappears.  Apparently, this conflicting desire of
the audience to simultaneously increase the average noise intensity
and to maintain synchronization leads to the sequence of appearing
and disappearing synchronized regimes.

These results indicate that the transition from random to
synchronized clapping is accompanied by a period doubling process.
Next we argue that in fact period doubling is a necessary condition
for synchronization.  To address this question, we investigated the
internal frequency of several individuals by controlled clapping
experiments. Individual students, isolated in a room, were instructed
to clap in the manner they usually do right after a good performance
(Mode I clapping), or during the rhythmic applause (Mode II
clapping).  As Fig.~1f shows, the frequencies of the Mode I and Mode
II clapping are clearly separated and the average period
doubles from Mode I to Mode II clapping.  Most important, however, we
find that the width of the frequency distribution and the relative
dispersion of the Mode II clapping is considerably smaller, a result
that is reproducible for a single individual as well (Fig.~1g).

These results indicate that after an initial asynchronous phase,
characterized by high frequency clapping (Mode I), the individuals
synchronize by eliminating every second beat, suddenly shifting to a
clapping mode with double period (Mode II) where dispersion is
smaller.  As shown by Winfree and Kuramoto, for a group of globally
coupled oscillators the necessary condition for synchronization is
that dispersion has to be smaller than a critical value
\cite{winfree,kuramoto}.
Consequently, period doubling emerges as a condition of
synchronization, since it leads to slower clapping modes during which
significantly smaller dispersion can be maintained.  Thus our
measurements offer a key insight into the mechanism of syncnronized
clapping:  during fast clapping synchronization is not possible
due to the large dispersion in the clapping frequencies.  After
period doubling, as Mode II clapping with small dispersion appears,
synchronization can be and is achieved. However, as the audience
gradually decreases the period to enhance the average noise intensity,
it gradually slips back to the fast clapping mode with larger
dispersion, destroying synchronization.

In summary, the individuals in the audience have to be aware that by
doubling their clapping period they can achieve synchronization,
which perhaps explains why in the smaller and culturally more
homogeneous Eastern European communities synchronized clapping is a
daily event, but it is only sporadically observed for the West and
U.S. In general, our results offer evidence of a novel route to
synchronization, not yet observed in physical or biological systems
\cite{bottani,glass,kuramoto,mirollo}.

\medskip
\medskip
\medskip

{\bf Z. N\'eda$^*$, E. Ravasz$^*$,
Y. Brechet$^{**}$, T. Vicsek$^{***}$,  A.-L. Barab\'asi$^{\&}$},

$^*$ Babe\c{s}-Bolyai University,  Department  of Theoretical Physics,
         str. Kog\u{a}lniceanu 1, RO-3400, Cluj-Napoca, Romania

$^{**}$ ENSEEG-LTPCM, INP-Grenoble
Saint-Martin D'Heres, CEDEX, France

$^{***}$  Department  of Biological Physics,
E\"otv\"os-Lor\'and University, Budapest, Hungary

$^{\&}$ Department of Physics, University of Notre Dame, IN 46556, USA

{\bf E-mail}: zneda@hera.ubbcluj.ro

\newpage


\newpage



\begin{figure}[htp]
\epsfig{figure=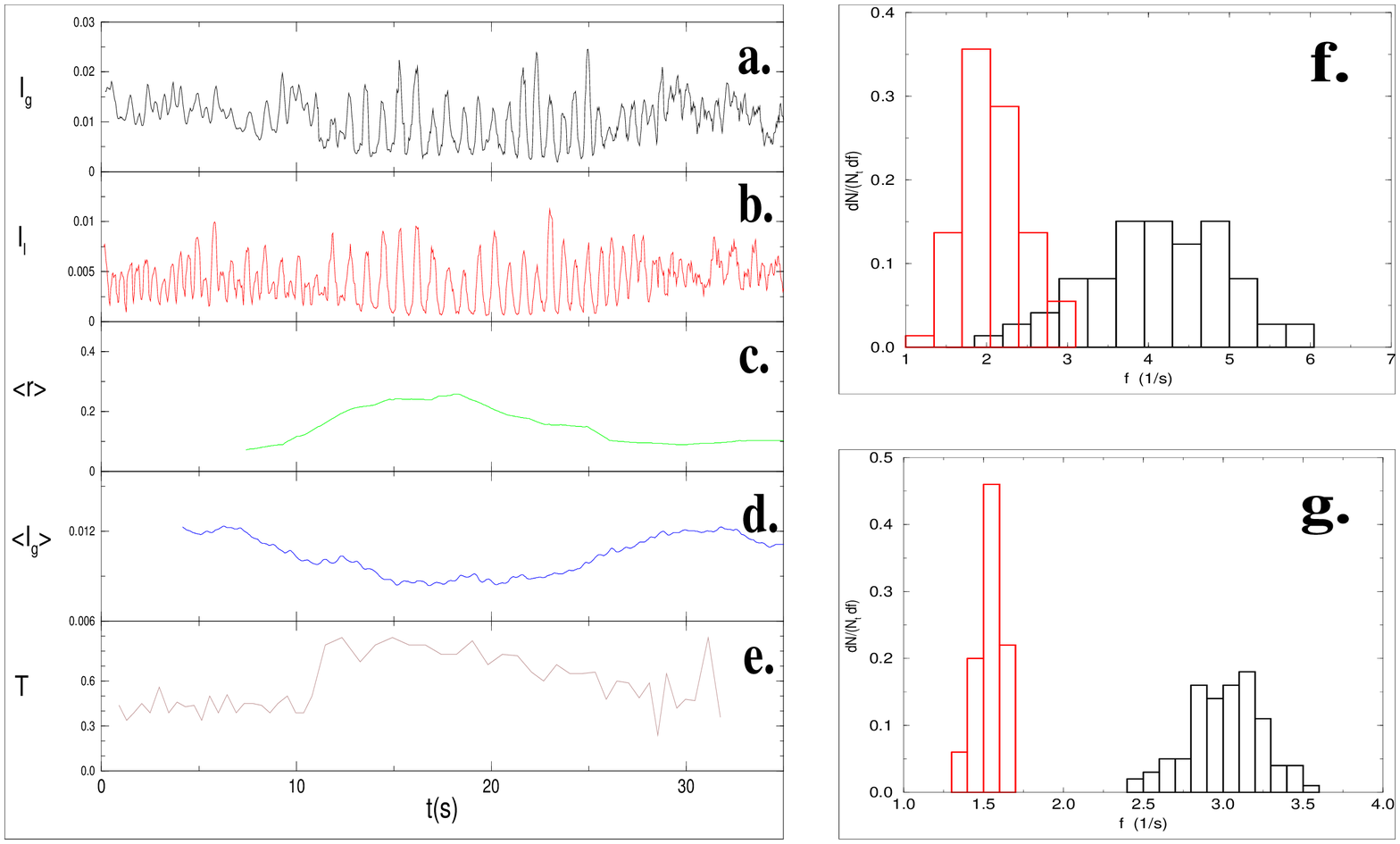, height=4.0in,width=6.0in,angle=-0}
\caption{Emergence of synchronization in clapping. (a) Global noise
intensity as a function of time.  The digitized data was squared and
moving average was performed over a window of size $0.2s$, much
smaller than the clapping period. The figure shows a
characteristic region with the appearance and disappearance of the
synchronized clapping. Over several performances we recorded 50
similar sequences of synchronized clapping. (b) Local noise intensity,
measured by a hidden microphone in the vicinity of a spectator.  (c)
Order parameter,
$r$, defined as the maximum of the normalized correlation between the
signal $c(t)$ and a harmonic function, $r=\max_{(T, \phi)}
\int_{t-T}^{t+T} c(t) \sin(2 \pi/T+\phi) dt / \int_{t-T}^{t+T} c(t) dt
$, where $\phi$ and $T$ span all possible values.
(d) Average noise intensity, obtained by taking a moving average
over a $3s$ window of the global noise intensity shown in (a).
(e) The clapping
period, defined as the intervals between the clearly distinguishable
maxima. (f) The normalized histogram of clapping frequencies measured
for 73 high school students (isolated from each other) for Mode I
(black) and Mode II (red) clapping.  (g) Normalized histogram for
Mode I and II clapping obtained for a single student, sampled 100
times over a one week period.}

\label{fig}
\end{figure}
\end{document}